\documentclass[final,5p,times,authoryear]{elsarticle}

\usepackage{amssymb}
\usepackage{aasmacros}
\usepackage{url}
\usepackage{xcolor}
\usepackage{graphicx}
\usepackage{float}
\usepackage{dblfloatfix}
\usepackage{amsmath}
\usepackage[linktoc=page]{hyperref}

\journal{Astronomy \& Computing}

\usepackage{caption}
\usepackage{subcaption}
\usepackage{booktabs}
\usepackage{siunitx}
\usepackage{placeins}
\usepackage{ragged2e}
\begin{document}

\begin{frontmatter}

\title{\LARGE\textbf{\emph{flashcurve}}\Large\textbf{: A machine-learning approach for the simple and fast generation\\ of adaptive-binning light curves with Fermi-LAT data}}

\author[inst1,inst2,inst5,inst6]{T. Glauch}
\author[inst1]{K. Tchiorniy}

\affiliation[inst1]{organization={Technische Universität München, Physik-Department},
            addressline={James-Frank-Str. 1}, 
            city={Garching bei München},
            postcode={D-85748},
            country={Germany}}

\affiliation[inst2]{organization={Institute for Advanced Study, Technische Universität München},
            addressline={Lichtenbergstrasse 2a}, 
            city={Garching bei München},
            postcode={D-85748}, 
            country={Germany}}

\affiliation[inst5]{organization={Heidelberg University, Institute of Environmental Physics},
            addressline={Im Neuenheimer Feld 229}, 
            city={Heidelberg},
            postcode={D-69120}, 
            country={Germany}}

\affiliation[inst6]{organization={Deutsches Zentrum für Luft- und Raumfahrt, Institut für Physik der Atmosphäre},                       city={Oberpfaffenhofen},
            country={Germany}}

\begin{abstract}
Gamma rays measured by the Fermi-LAT satellite tell us a lot about the processes taking place in high-energetic astrophysical objects. The fluxes coming from these objects are, however, extremely variable. Hence, gamma-ray light curves optimally use adaptive bin sizes in order to retrieve most information about the source dynamics and to combine gamma-ray observations in a multi-messenger perspective. However, standard adaptive binning approaches are slow, expensive and inaccurate in highly populated regions. Here, we present a novel, powerful, deep-learning-based approach to estimate the necessary time windows for adaptive binning light curves in Fermi-LAT data using raw photon data. The approach is shown to be fast and accurate. It can also be seen as a prototype to train machine-learning models for adaptive binning light curves for other astrophysical messengers. 

\end{abstract}


\end{frontmatter}

\section{Introduction}
Gamma rays are photons at the high-energy end of the electromagnetic spectrum, with energies starting from a few hundred keV and going up to ultra-high energies in the TeV and PeV ranges. They are produced in various processes like nuclear decays and the collision of high energy particles \citep{Gaisser_Engel_Resconi_2016}.
Hence, they are a unique way to study the high-energy phenomena in our universe. 

The Fermi-Large Area Telescope (LAT) is a satellite that studies gamma rays from Galactic and extragalactic environments in the energy range between 100 MeV and a few hundred GeV. Several thousand gamma-ray point sources have already been detected by Fermi-LAT in the extragalactic sky \citep{ballet2024fermilargeareatelescope}.

The sources at high Galactic latitudes are mostly blazars - active galactic nuclei with a jet of high-energy particles pointing towards the observer - but also starburst- and radio galaxies \citep{Ajello:2015mfa}. It has been shown that the gamma-ray fluxes from those objects are, in most cases, highly variable \citep{SBARRATO2011998}. In order to catch the dynamics of the processes in those sources, it is necessary to calculate time-dependent light curves. This gets even more important in the emerging field of multi-messenger astronomy, which tries to combine (time-dependent) measurements of different particles (e.g. photons, neutrinos, protons) at different energies (for sub-eV to PeV) to get a comprehensive picture of the processes happening within our universe. Many of those studies already use such approaches or could greatly profit from doing so \citep{10.1093/mnras/sty1852,10.1093/mnras/staa2082}. 

Due to the high variability and varying strength of gamma-ray sources, it is challenging to choose an adequate time width for each light curve bin such that the integration time is long enough to catch the signal but short enough to catch relevant features. For a long time, the preferred method for choosing the (adaptive) length of the time bins was the so-called adaptive binning approach \citep{Lott_2012} that aims to produce light curves with bins having a constant significance or flux uncertainty $\Delta_0$. Numerically, this requires to solve the equation
\begin{equation}
    \sigma_f(T_0, T_1, \bar{\gamma}, \bar{F} ) = \Delta_0
\end{equation}
for the ending time of a time interval $T_1$ giving some starting time $T_0$ and some function $\sigma_f$. $\bar{\gamma}$ and $\bar{F}$ are the average flux and photon spectral index over the interval [$T_0$, $T_1$], respectively. However, the (sequential) search for the time bins fulfilling this equation is extremely expensive, so the computation of $\sigma_f$ requires approximations instead of a full likelihood model. Nevertheless, with traditional methods, a single adaptive light curve often takes several hours, if not days, to compute. This paper presents a novel approach for the (fast) generation of adaptive binning for light curves that uses a machine-learning approach to approximate the Fermi likelihood function.

In recent years, machine-learning-based methods have become quite popular in astronomy as they are quick in the calculation and provide a high accuracy \citep{fluke2020surveying, sen2022astronomical, GLAUCH2022100646,2019arXiv190808763K}. As a result, our new tool\footnote{Available at \url{https://github.com/kristiantcho/flashcurve}}, \textit{flashcurve}, allows large-scale generation of adaptive-binning for gamma-ray light curves as well as quick follow-up studies in astronomical real-time programs. 

The paper is structured as follows. In section 2, we present the dataset of light curves that we use to train the machine-learning estimator; in section 3, we explain its structure and the training process; and in section 4, we show how it can be used. We conclude with a summary in section 5. 
\label{sec:sample1}

\section{Data sample of gamma-ray light curves}
\label{sec:data_sample}

\subsection{Fermi-LAT LCR \& analysis}

The Fermi-LAT Light Curve Repository (LCR) \\\noindent\citep{Abdollahi_2023} is a database of light curves including 1525 sources from the 10-year Fermi-LAT point source 4FGL-DR2 catalogue \citep{ballet2020fermilargeareatelescope}. This database and the corresponding raw photon data downloaded from the Fermi LAT data server were used to train the machine-learning estimator model presented here.

The sources included in the LCR database have a variability index greater than 21.67, indicating the average fractional flux variability (dF/F) measured on timescales of 1 year. Having a variability over this value for over 12 intervals means that these sources are estimated to have a \textless 1\% chance of being a steady source \citep{Abdollahi_2020}. The resulting source selection comprises blazars, further classified as flat spectrum radio quasars, BL Lacs, and blazar candidates of unknown type.

Each source in the catalogue has light curves with three different fixed time-bin lengths: 3 days, 1 week, and 1 month. This was done for over 14 years of data, from 2008 to the end of 2023, which accumulated to over 3.7 million individual time bins. 

The gamma-ray flux of each time bin was estimated using a maximum likelihood analysis as described in \\\noindent\citet{Abdo_2009}. The signal hypothesis of the likelihood function includes a power-law point source flux $\phi$ on top of a diffuse gamma-ray background and the known gamma-ray sources in the region. Free parameters are the flux normalization $\phi_0,$ and the spectral index $\gamma$ i.e.,
\begin{equation}
    \phi(E_{\gamma};\gamma, \phi_0)  = \phi_0 \times (E_{\gamma}/E_0)^{-\gamma} 
\end{equation}
with the photon energy $E_{\gamma}$.
The test statistic (TS) quantity is defined as
\begin{equation}
    \mathcal{TS} = -2\ln{\left(L_{\text{max,0}}/L_{\text{max,1}}\right)}
    \label{eq:TS}
\end{equation}
% \vspace{0em}
where $L_{\text{max,0}}$ is the maximum likelihood value for a model without an additional source (the null hypothesis) and $L_{\text{max,1}}$ is the maximum likelihood value for the signal hypothesis.

In order to maximise the signal and background likelihood, the observed energy and direction of the photons are used with Fermi's instrument response functions. The exact location of the source is found through a grid search for the maximum test-statistic value. The analysis settings for the Fermi-LAT LCR analysis can be found in Table \ref{T:LCR}. Our machine-learning algorithm uses the same data settings regarding energy range, event selection, instrument response function and acceptance cone.
For Fermi point sources analyses, the TS is distributed approximately as a $\chi^2$ distribution \citep{Abdo_2009}; thus, the square root of the TS is approximately equal to the detection significance for a given source \citep{Wilks:1938dza}. This TS will be the target variable for our machine-learning algorithm using the raw photon data as input. The LCR analysis rejects the background hypothesis only when $\mathcal{TS} \geq 4$.

\begin{table}[H]
\fontsize{8.5}{11}\selectfont
\centering
\begin{tabular}{|l|l|}
\hline
\textbf{Time Bins} & 3 day, 7 day, 30 day \\ \hline
\textbf{Energy Range} & 100 MeV - 100 GeV  \\ \hline
\textbf{Event Selection} & P8R3\_SOURCE  \\ \hline
\textbf{Instrument Response Function} & P8R3\_SOURCE\_V2 \\ \hline
\textbf{Acceptance Cone (ROI)} & 12 deg (radius) \\ \hline
\textbf{Zenith Angle Cut (zmax)} & 90 deg \\ \hline
\textbf{Fit optimizer} & MINUIT \\ \hline
\textbf{Interstellar Emission Model} & gll\_iem\_v07.fits \\ \hline
\textbf{Isotropic Spectral Template}	& iso\_P8R3\_SOURCE\_V3\_v1 \\ \hline
\textbf{4FGL-DR2 Catalog} & gll\_psc\_v27.fit \\ \hline
\end{tabular}
\caption{Analysis settings for the the Fermi-LAT LCR analysis.}
\label{T:LCR}
\end{table}

To obtain the LCR light curve data conveniently, we used the Python API script \emph{pyLCR}\footnote{\url{https://github.com/dankocevski/pyLCR}, \raggedright last accessed: 01.04.2024}, which was used to get the exact time bins and corresponding TS for each source. We include time bins with a minimum TS of 0. By including time bins with TS $<$ 4, we aim to train the estimator so that it possibly recognises the background. For these time bins, however, only upper limits for fluxes are provided. 

\subsection{Data sample cleaning}

The LCR data source comes with some caveats. One caveat was that some results may have come from non-convergent \\\noindent analyses. Thus, we removed these samples from the final data sample selection by checking for nonzero MINUIT \citep{James:1975dr} return codes. Some data points had negative TS values, resulting from non-convergent fits. Those data points were also removed.

\begin{figure}[h!]
    \centering
    \includegraphics[scale=0.2]{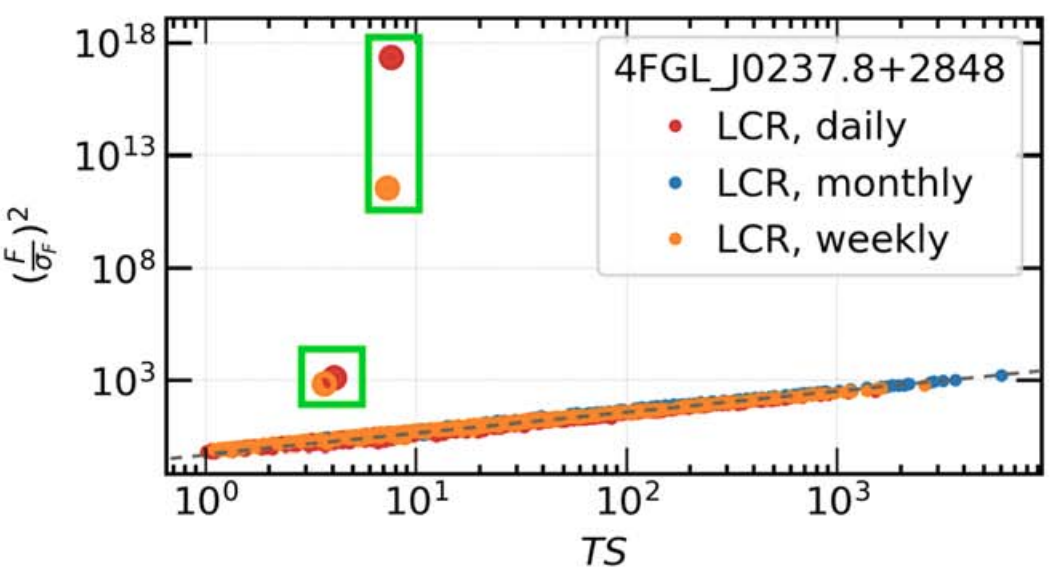}
    \caption{An example of an LCR 'sanity check' for time bins with analyses which may have failed. If a flux-to-flux uncertainty ratio within a time bin wildly deviates from an approximately linear behaviour over the square root of the test statistic, this time bin should not be used in high-level analysis. Plot taken from \citet{Abdollahi_2023}.}
    \label{fig:LCR_sanity}
\end{figure}

\justifying
Apart from this, the LCR usage notes\footnote{\url{https://fermi.gsfc.nasa.gov/ssc/data/access/lat/LightCurveRepository/about.html}, last accessed: 01.10.2024} recommend performing `sanity checks' to check whether the flux uncertainty to flux to ratio, $\sigma_{F}\allowbreak/F$, behaves approximately linearly to the square root of the TS, as well as to check whether fluxes are deviating quite strongly from the total distribution of fluxes for a given source. Figure \ref{fig:LCR_sanity} taken from \citet{Abdollahi_2023} shows an example of data which deviates from the linear approximation. However, we found no deviations in our selected sources when performing such a sanity check with the data. 

\begin{figure}[h!]
    \centering
    \includegraphics[scale=0.6]{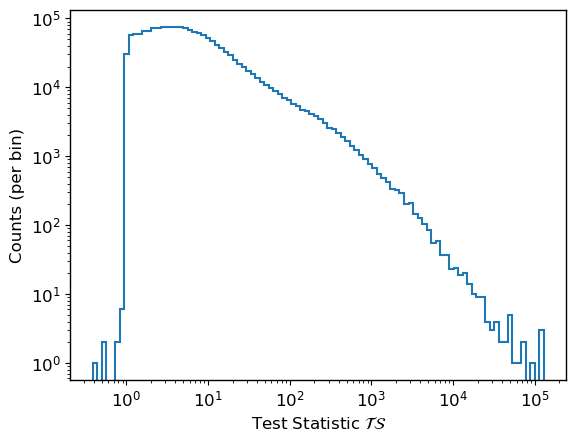}
    \caption{Binned distribution of test statistic from the Fermi LCR time bins which were to be used for training the estimator model.}
    \label{fig:TS_dis}
\end{figure}

Finally, we removed sources within 10 degrees of the Galactic plane to avoid confusion with the Galactic background. \\Overall, 163 sources. The LCR database also includes a few extended sources, which we excluded. All of the aforementioned filters amount to a total of 1362 sources and around 1.5 million time bins that could be used to train the estimator model. Figure \ref{fig:TS_dis} shows the distribution of the TS for all the selected time bins.

\section{Machine-learning-based adaptive light curve}
\label{sec:ml_algorithm}
\subsection{General idea}

Artificial neural networks have made huge strides in the last few years, especially in computer vision. Many techniques and architectures have been developed to increase the prediction quality of image recognition, both in categorical and regression estimation \citep{c-vision} \citep{cnn_vision}. Similarly, we use the spatial and energy information of the time-integrated Fermi-LAT observations to estimate the detection significance.

For training data, we use the previously discussed Fermi-LAT LCR dataset. Although this dataset only covers light \\curves with three different time-bin lengths, it covers an large range of TS values due to the sources' variability. Moreover, sources are distributed isotropically in the sky, avoiding directional bias in the training.

\subsection{Data preparation}
\label{ss:image}
Overall, our training dataset contains 1.5 million time bins. We generated neural network input images for each time bin by binning the photon counts in right ascension (RA), declination (Dec), energy, and time. 

Similar approaches have been used in \citet{Caron_2018}, which binned photon counts in right ascension, declination and several energy bins as image channels instead of a single (broad range) channel. 

\begin{figure}[h]
    \centering
    \includegraphics[scale=0.36]{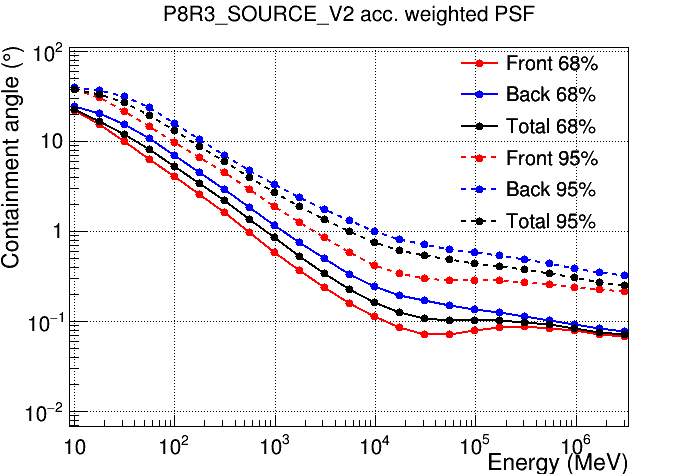}
    \caption{68\% and 95\% containment angles of the acceptance weighted PSF for both the front/back PSF event types for the Fermi-LAT P8R3\_SOURCE\_V2 instrument response functions. Plot taken from \sloppy\url{https://www.slac.stanford.edu/exp/glast/groups/canda/archive/pass8r3v2/lat_Performance.htm}, last accessed: 01.03.2024}
    \label{fig:passv2}
\end{figure}

\begin{figure*}[h!]
    \centering
    \includegraphics[scale=0.83]{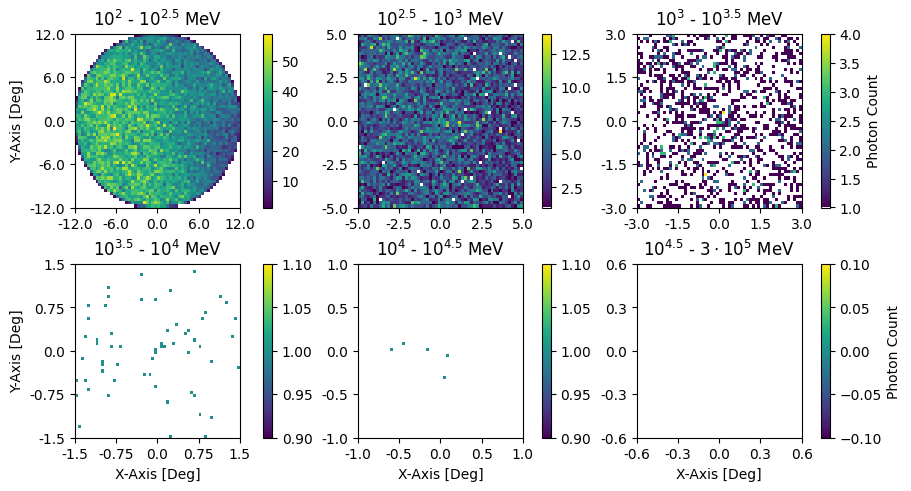}
    \caption{Example neural network input image for 4FGL J0319.8+4130 used to train the estimator model. The total input has six (energy bin) channels with photon counts binned in angular distance for a tangential plane around the source. The colour bars indicate counts in each bin. Each panel has counts of photon events within a specified energy range (energy bin). The source depicted here has a TS of 569.12.}
    \label{fig:medts}
\end{figure*}

The generation of the input images for the neural network has two steps. Firstly, the photons are rotated into a spherical coordinate system in which the source is located at a pole. We do this to standardise the angular distances of each photon event from the source. Secondly, the photons are binned into four dimensions: time, energy, and the two tangential dimensions of the rotated spherical coordinates, which we denote as X and Y in the flattened image (see Figure \ref{fig:medts}). In order to avoid creating an exceedingly large feature space, only six energy bin channels were used. Thus, similar to energy bins typically used in Fermi-LAT spectral energy distributions, energy was binned in half powers of 10 MeV, ($10^2$ MeV, $10^{2.5}$ MeV, $10^{3}$ MeV, $ 10^{3.5}$ MeV, $10^{4}$ MeV, $10^{4.5}$ MeV, and  $3\cdot 10^5$ MeV). Since the point spread function of Fermi-LAT decreases with energy (Figure \ref{fig:passv2} ), we decrease the range of the X-Y histogram with increasing energy \citep{atwood2013pass8realizationfermilat} to 12 deg, 5 deg, 3 deg, 1.5 deg, 1 deg, and 0.6 deg following the 95\% quantile of the Fermi-LAT point spread function. In each dimension, 56 bins are used. 

\begin{figure*}[h!]
    \centering
    \includegraphics[scale=0.83]{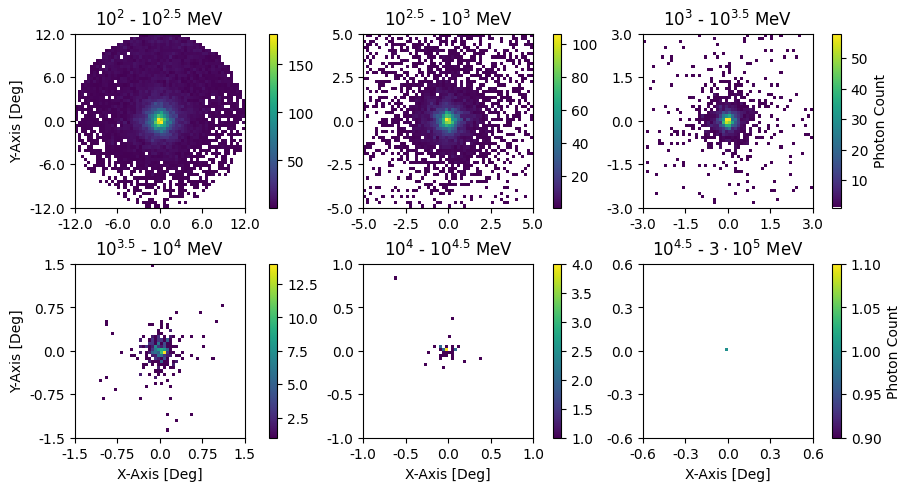}
    \caption{Same as the figure before, but using data from the source 4FGL J2253.9+1609, which, in the given time window, had a TS of 129479.46, making it the most significant time bin in the data sample.}
    \label{fig:hights}
\end{figure*}

Figures \ref{fig:medts} and \ref{fig:hights} illustrate the neural network inputs generated with this procedure. The first image is from \\4FGL J0319.8+4130, and the second comes from the source 4FGL J2253.9+1609, which had the highest significance time bin of all samples. Both images depict time bins of 1 month with significant gamma emission (TS of $\sim570$ and $\sim1.3\cdot10^5$ corresponding to significances of $\sim25\sigma$ and $\sim360\sigma$ respectively). 

Figure \ref{fig:hights} visually shows clear emission from the source in all energy channels, while in Figure \ref{fig:medts}, it is less visible to the human eye. Especially at low energies, photons from a nearby source are dominating the image. The picture is clearer at higher energies, contributing most to the overall significance. 

\subsection{The artificial neural network - architecture \& training}
\label{ss:train}

\subsubsection{Convolutional \& residual neural networks}

Convolutional neural networks (CNN) are a class of deep neural networks (DNN) that are designed to be trained to make predictions using visual data or images \citep{cnn}. A CNN generally takes an N-dimensional grid input and uses convolutional layers and non-linear activation functions to map the input to an M-dimensional output. The input can be, for example, a multichannel image with dimensions ($w,h,c$), $w, h$ being the width and height of the image in terms of the number of pixels and then $c$ the number of channels. In the case of a colour image, we have three channels: red, green, and blue. In our case, we have six energy bins. The output could be a set of predictions for different image class options or, as in our case, a single number (regression problem).

\begin{figure}[H]
    \vspace*{-0.5cm}
    \centering
    \includegraphics[scale=1]{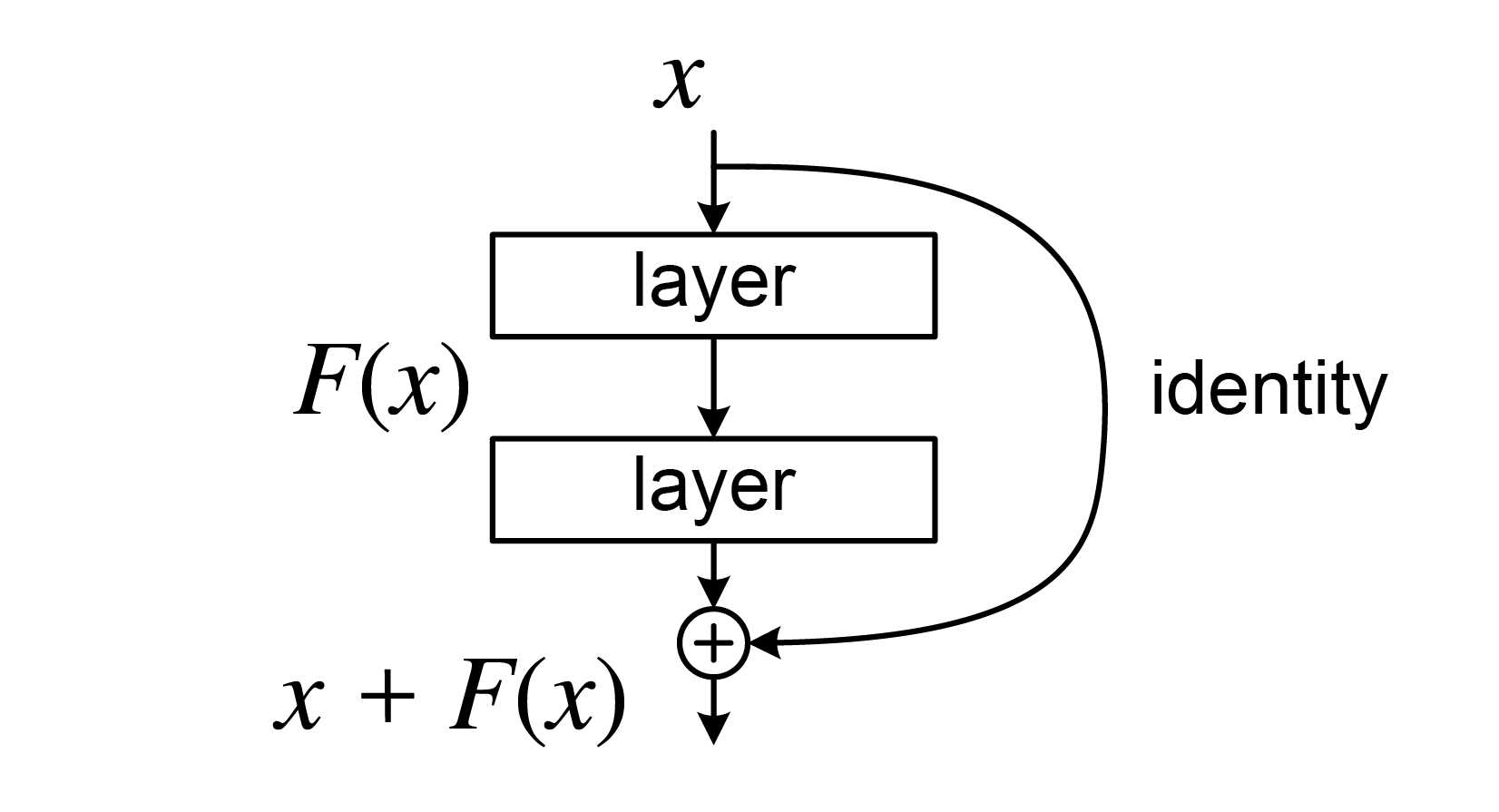}
    \caption{The basic working principle of a residual block used in ResNets, which consists of an input $x$ having transformations $F(\,)$ applied to it, whilst having an identity mapping added onto its output in the form of a 'skip connection': $x + F(x)$. Image taken from Wikimedia Commons \url{https://commons.wikimedia.org/wiki/File:ResBlock.png}, last accessed: 01.10.2024.}
    \label{fig:resblock}
\end{figure}

Additional improvements in network performance can be achieved using residual neural networks (ResNets) \citep{he2015deepresiduallearningimage}. Instead of learning independent features in each layer, a residual neural layer only learns corrections to the previously learnt features. This can be implemented by including an identity connection between subsequent layers, as illustrated in Figure \ref{fig:resblock}. 

A ResNet consists of a sequence of multiple residual blocks. This is because stacking more layers in a typical DNN could lead to a swift reduction in training accuracy, often referred to as the 'degradation' problem \citep{he2015deepresiduallearningimage}. If a DNN is able to reach a maximal training accuracy with the initial layers (or some other sub-combination of layers inside the architecture), then these skip connections would ensure that only the additional unnecessary layers are skipped and that only the identity mapping is kept. ResNets have been shown to increase training stability and, therefore, accuracy \citep{resnetperf}.

\subsubsection{Flashcurve architecture \& training}

\begin{figure}[h!]
    \centering
    \includegraphics[scale=0.27]{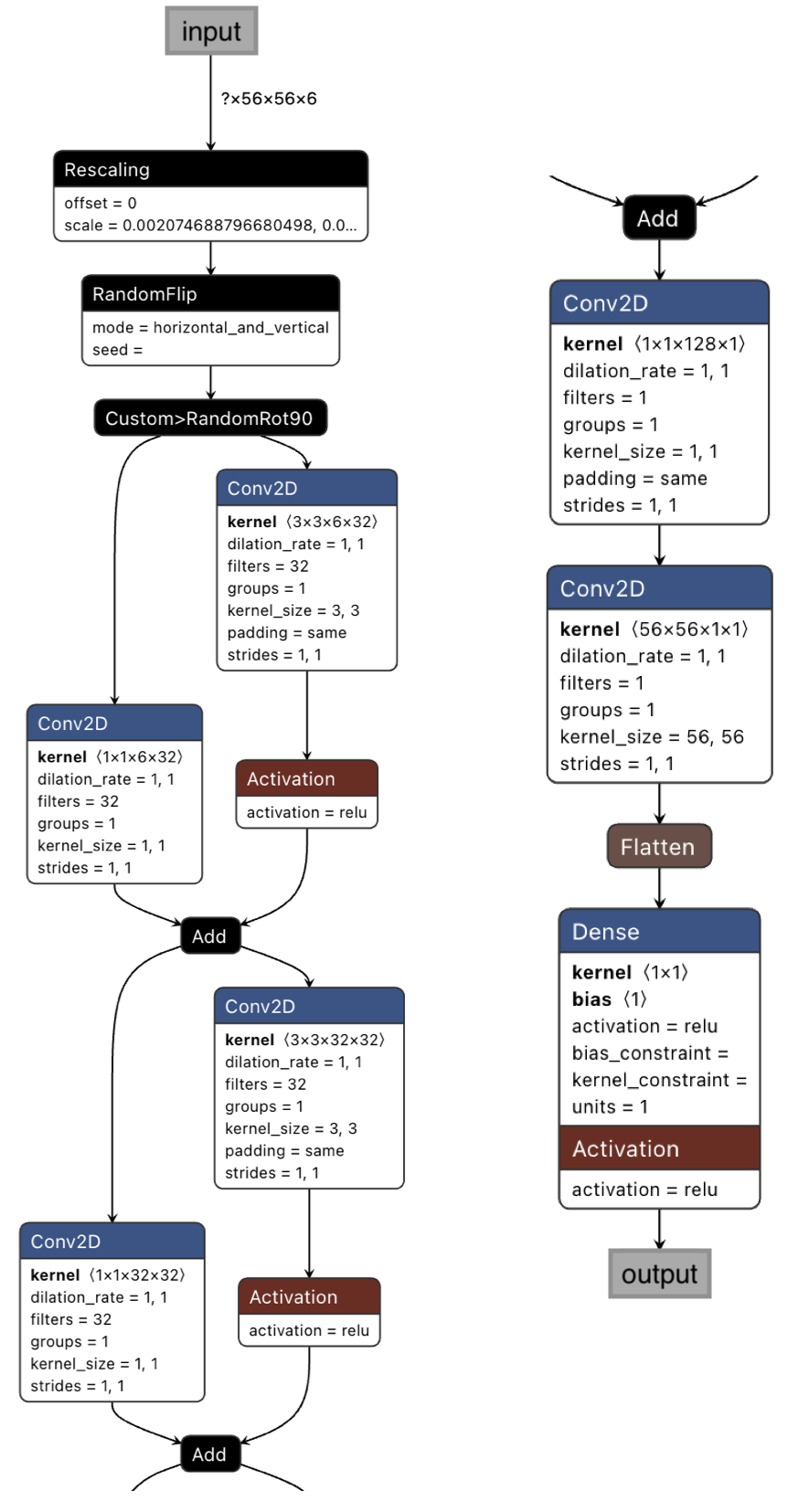}
    \caption{\emph{Left:} First layers of the \textit{flashcurve} estimator model. All layers between two consecutive 'Add' layers comprise a residual block. Included here are also the 'Rescaling' preprocessing layer, which normalises the images, and the 'Random Flip' and 'Random Rotation' layers, which are only used during training to horizontally flip or rotate the images in 90-degree steps. \\
    \emph{Right:} Last layers after the residual blocks, including a 'Dense' layer with a \emph{RELU} activation, which acts as a linear output for the TS estimation.}
    \label{fig:model-arch}
    % \vspace*{-0.5cm}
\end{figure}

 Figure \ref{fig:model-arch} depicts the network architecture of \textit{flashcurve}. The model consists mainly of the previously mentioned residual \\blocks, each with a 3x3 convolutional layer with a stride of 1 and a filter count of 32, 64, or 128. At the end of each layer, we use the \emph{RELU} function as a non-linear activation layer.

The identity mapping is performed with a 1x1 convolutional layer, which ensures that the number of channels in both layers matches when added at the end of the residual block. 

Some preprocessing steps are performed in the input layer: First, a rescaling layer normalises the counts in each energy channel using the maximum values found in the training data set. In addition, we used a random flip layer as well as a rotation layer, which rotates the images in random multiples of 90 degrees. Those data augmentations increase the size of the training dataset by a factor of 8, allowing a better model generalization\citep{perez2017effectivenessdataaugmentationimage}.

In the network's last layer, the network features are mapped to a single number, the test statistic, see Figure \ref{fig:model-arch}. The final linear layer has another \emph{RELU} activation to ensure that the TS value is a strictly positive number. 

We trained the model with a batch size of 256 using the \\\emph{Tensorflow} python package, the \emph{Keras} API, the \emph{ADAM} \\\citep{kingma2017adammethodstochasticoptimization} optimiser, and a mean-squared error (MSE) loss function.
Table \ref{tab:model_sum} summarises our model's total number of residual blocks. This combination of layers was chosen such that the number of parameters is one order of magnitude smaller than the total number of samples. In total, the network has $\sim\,$400000 trainable parameters.

\begin{table}[H]

\fontsize{8}{9.5}\selectfont
\centering
\begin{tabular}{|l|l|}
\hline
\textbf{\# of Res. Blocks} & \textbf{\# of filters} \\ \hline
15 & 32  \\ \hline
5 & 64  \\ \hline
1 & 128 \\ \hline

\end{tabular}
\caption{Ordered list of residual blocks and the corresponding number of filters used in the convolutional layer within each block used for the final estimator model.}
\label{tab:model_sum}
\end{table}

\subsection{Validation \& performance}

\begin{figure}[h!]
    \centering
    \includegraphics[scale=0.38]{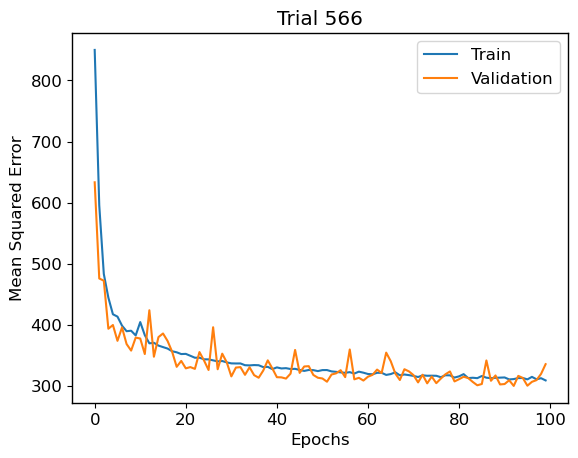}
    \caption{Learning curve displaying the output of the mean squared error loss function during the training of the best estimator model at different epochs for both the validation and training set.}
    \label{fig:lrn-c}
\end{figure}

In order to test the model performance and its ability to generalise, we split the dataset into three sets: 80\% was used as a training set, 10\% as a validation set, and the remaining 10\% as a test set, which would remain unseen until the model completed training. The validation set is used during training to test the model's performance after each epoch. It is also used to select the final model based on the epoch at which the model produced the best validation loss. Figure \ref{fig:lrn-c} shows the output of the loss function over all the epochs for our best-performing model.

\begin{figure}[h!]
    \centering
    \includegraphics[scale=0.37]{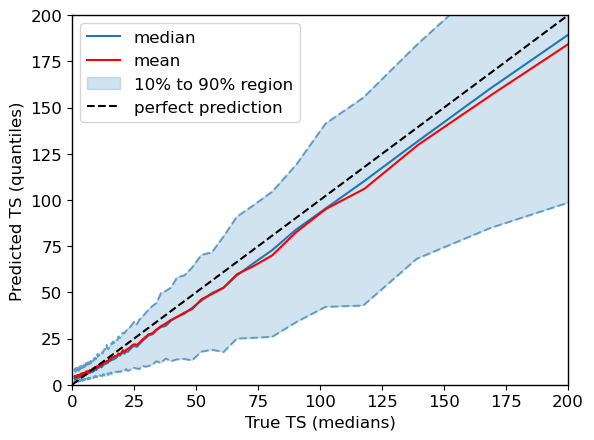}
    \caption{Predicted TS from an unseen test set of images vs. the true TS. To simplify the plot, mean, median, and the 10\% to 90\% quantile region of binned values are taken and plotted against the medians of the binned true TS.}
    \label{fig:ts_per}
\end{figure}

Figure \ref{fig:ts_per} shows the model's performance on the unseen test set, which achieved approximately constant relative uncertainty across the entire TS space. 

\begin{figure}[h!]
    \centering
    \includegraphics[scale=0.39]{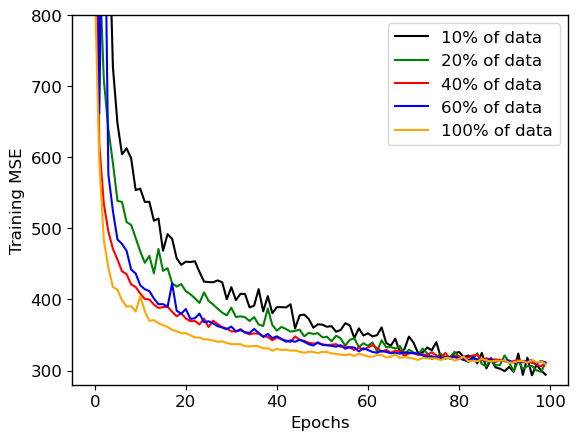}
    \caption{Learning curve displaying the output of the mean squared error (MSE) loss function during the training of an estimator model at different epochs for different subsets of the main training set.}
    \label{fig:loss_test}
\end{figure}

\begin{figure}[h!]
    \centering
    \includegraphics[scale=0.39]{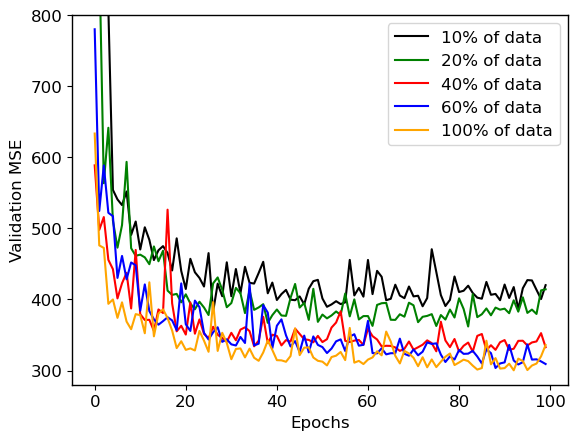}
    \caption{Learning curve displaying the output of the mean squared error (MSE) loss function during the training of an estimator model at different epochs, with different subsets of the main training set, and testing then on the main validation set.}
    \label{fig:val_test}
    \vspace{-0.2cm}
\end{figure}

To evaluate how the size of the training dataset impacts the model performance, we rerun the training procedure with different subsets of the training dataset, all while still keeping the same validation and test set. The results are shown in Figures \ref{fig:loss_test} and \ref{fig:val_test}. The losses on the training set show that more data leads to a more stable training; meanwhile, the validation losses indicate that lower losses are achieved with more data. This indicates that the amount of training data available currently limits our model.

\subsection{Investigation of bad predictions}

\begin{figure}[h!]
    \centering
    \includegraphics[scale=0.4]{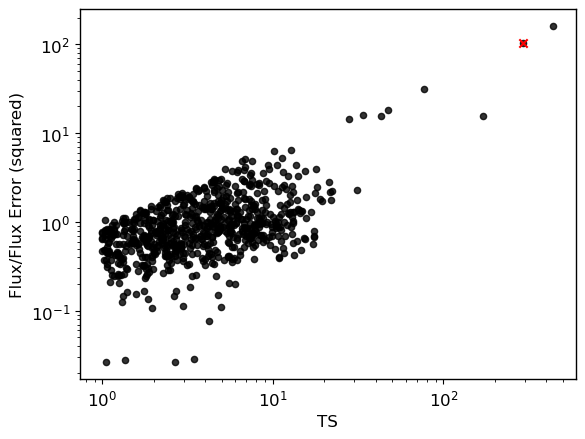}
    \caption{Set of TS vs. their corresponding squared ratio of flux/flux uncertainty from the Fermi-LCR for a 3-day binned lightcurve of 4FGL J1234.0-5735. The red cross indicates a TS, which the estimator had vastly under-predicted. However, it does not appear very clearly here to be an outlier.}
    \label{fig:btsr}
\end{figure}

\begin{figure}[h!]
    \centering
    \includegraphics[scale=0.4]{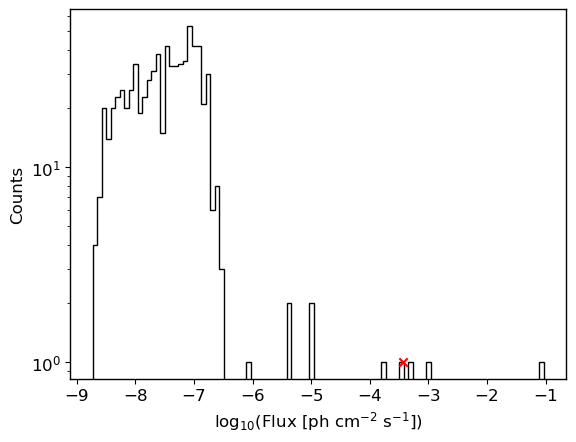}
    \caption{Log-binned counts of flux data of 4FGL J1234.0-5735, see \ref{fig:btsr} for more details. The badly predicted TS (red cross) appears to be an outlier from the distribution.}
    \label{fig:btsf}
\end{figure}

\begin{figure}[h!]
    \centering
    \includegraphics[scale=0.4]{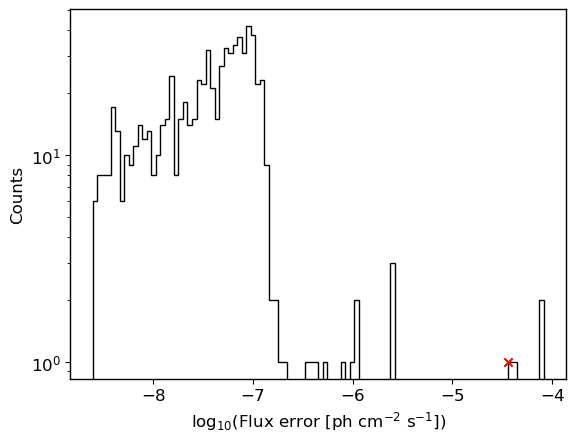}
    \caption{Log-binned counts of flux uncertainty data of 4FGL J1234.0-5735, see \ref{fig:btsr} for more details. The badly predicted TS (red cross) also appears to be an outlier from the distribution.}
    \label{fig:btsfe}
\end{figure}

We find two types of bad predictions from the test dataset: the model predicts a low TS, whereas the true TS is orders of magnitude higher, or vice versa. The former case was investigated by performing the checks recommended by \citet{Abdollahi_2023} and then subsequently running \emph{Fermipy} to cross-check the TS. One such case occurred with the source 4FGL J1234.0-5735, in which a TS of $\sim$2 was predicted, but the Fermi-LCR stated a TS of $\sim$300. Figures \ref{fig:btsr} - \ref{fig:btsfe} show the recommended checks performed.

\begin{figure*}[h!]
    \centering
    \includegraphics[scale=0.79]{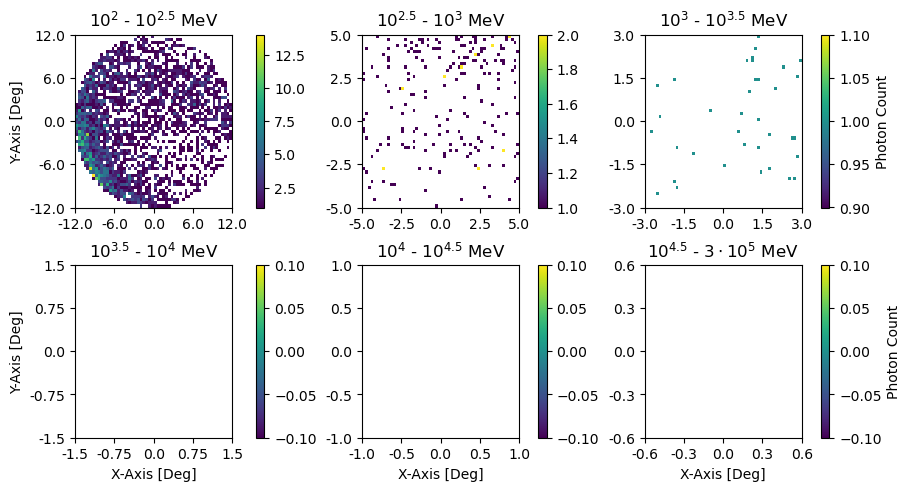}
    \caption{Image of binned photon event counts for 4FGL J1234.0-5735 in a 3-day period, corresponding to a TS $\sim$300, for which the estimator model under-predicted a TS of only $\sim$2. See Figure  \ref{fig:medts} for more information.}
    \label{fig:badts}
\end{figure*}

The badly predicted TS does not appear to be an outlier when looking at Figure \ref{fig:btsr}, which, according to the Fermi-LCR, would be the main reason to investigate further. However, looking at the distributions of flux and flux uncertainty, this data point exhibits a sudden burst of flux compared to the typical lower activity of the source. Figure \ref{fig:badts} depicts the respective image with which the model under-predicted the TS. 

\begin{figure*}[h!]
    \centering
    \includegraphics[scale=0.79]{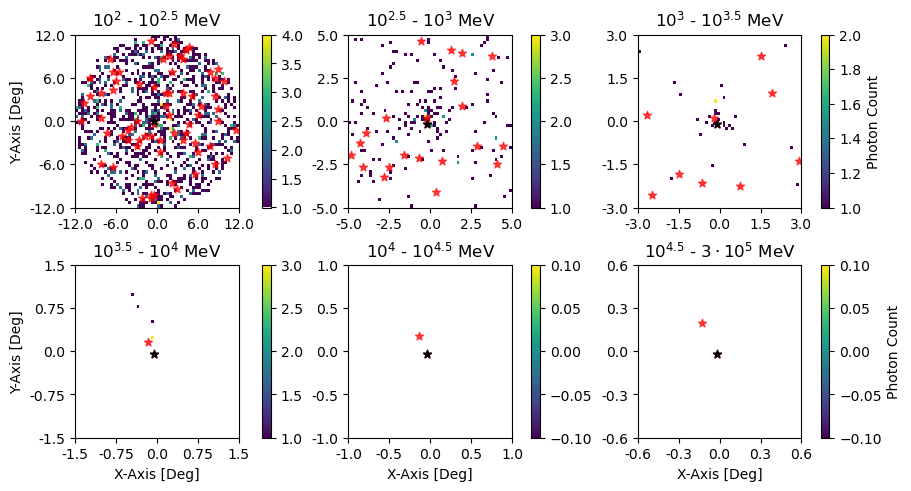}
    \caption{Same as the figure above for  4FGL J1311.0+3233 and a time bin of 1 week. Fermi-LCR reports a TS of 3.63, whereas the estimator model predicted $\sim$300. Additionally, overlaid on this image as red stars are the (relative) positions of nearby sources; the black star is 4FGL J1311.0+3233. It is apparent in this image that there is a high amount of activity close to the central source.}
    \label{fig:nearbyts}
    \vspace*{0.2cm}
\end{figure*}

Finally, the time bin (3-days centred at 673617601 MET) was rerun independently through \emph{Fermipy} and along with the 3-day bin before and after the time bin in question, all of which came out with extremely low flux ($O(10^{-10})$) and extremely low TS ($O(10^{-5})$). Thus, we concluded that this was an outlier in the LCR database that was missed during the initial data cleaning. 

However, judging by the flux and flux uncertainty distributions, one can see that the expected amount of outliers per source could be in the order of $10$, which would mean that the total number of expected outliers in the entire data set would be in the order of $10^4$, two orders of magnitude less than the total number of data-points. Hence, the impact of these outliers on the model is expected to be small. For the future, we plan to create more stringent filters for the data cleansing, for example, by filtering out data points whose flux or flux uncertainty sit outside 5\% and 95\% quantiles as in Figures \ref{fig:btsf} and \ref{fig:btsfe}. 

In the opposite case, there were numerous over-predictions from the model, one example of such being a prediction of $\sim$300 for a true TS of $\sim$3.6. One of the sources in question was 4FGL J1311.0+3233 for a time bin centred at 242308801 MET with a duration of 1 week. Figure \ref{fig:nearbyts} shows the image with photon counts along with the relative positions of nearby sources overlaid. We see that there is a considerable amount of gamma-ray sources close to 4FGL J1311.0+3233, even at relatively small angular distances, the most prominent being OP 313, with a source strength (for photon flux $> 1$ GeV) $\sim\,$4 times larger than 4FGL J1311.0+3233. This causes source confusion, something which is also not handled well by \citet{Lott_2012}. On the other hand, the complete Fermi-LAT analysis correctly accounts for the fluxes from all surrounding sources.
To handle source confusion better, in the future, we plan to explore a model which additionally incorporates these nearby source positions and fluxes, perhaps as a seventh layer on top of the current image format or overlaid on each layer as in Figure \ref{fig:nearbyts}. 

\subsection{Time bin search algorithm}

\begin{figure}[h!]
    \centering
    \includegraphics[scale=0.34]{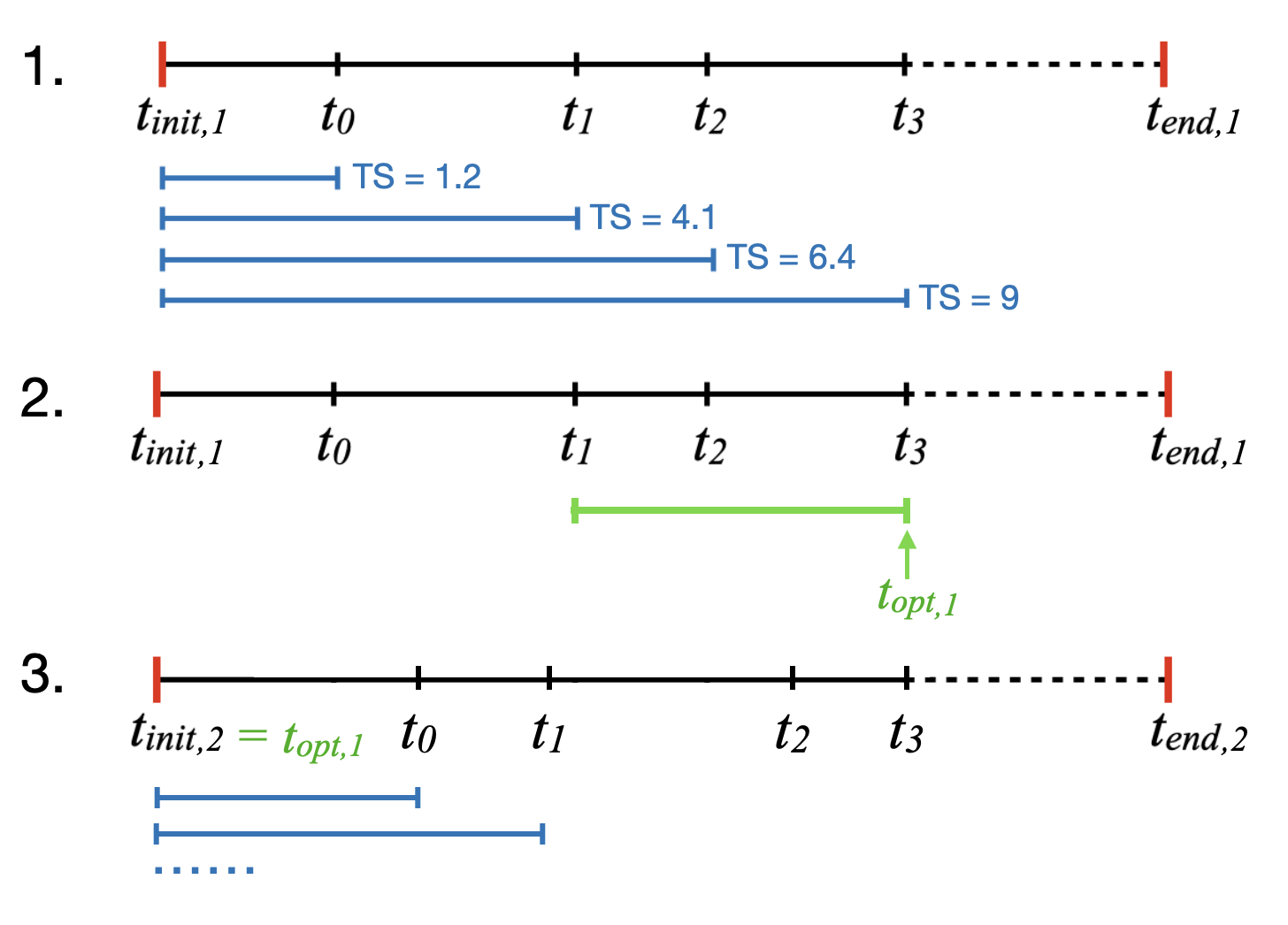}
    \caption{Visualisation of the steps of the time bin search algorithm with a TS target range of [4,9]. }
    \label{fig:tbins}
\end{figure}

The machine-learning-based TS estimation is central to our new adaptive light curve approach. We also require an algorithm that partitions the entire data range into possible time windows. Different methodological and numerical approaches could be considered here, independent of the machine-learning estimator. Our choice is depicted in Figure \ref{fig:tbins} and consists of the following steps:

\begin{enumerate}
    \item A time window (red interval) is searched with sub-time intervals (blue intervals) based on the time stamps of selected photon events $t_n$. Each interval uses the time window's lower bound ($t_{init}$) as its own lower bound. To speed up the search, only photons with a relatively large impact on the TS are considered for calculating the intermediate's time stamps $t_n$. The TS is predicted using the machine-learning estimator for each resulting time interval. \\
    \item The time stamps whose intervals produce TS within a TS target range (here [4,9]) are within the green interval. The \textbf{last} largest time window within the green interval is then used as the end time $t_{opt,1}$ for the light curve. If the TS are below the target range, the time window gets extended, and further events are processed. \\
    \item The search for more time bins continues repeatedly with a new time window from each $t_{opt}$ onwards until all photons are parsed through. The last bin will, therefore, not necessarily have a TS above the threshold.
\end{enumerate}

\begin{figure}[h!]
    \centering
    \includegraphics[scale=0.34]{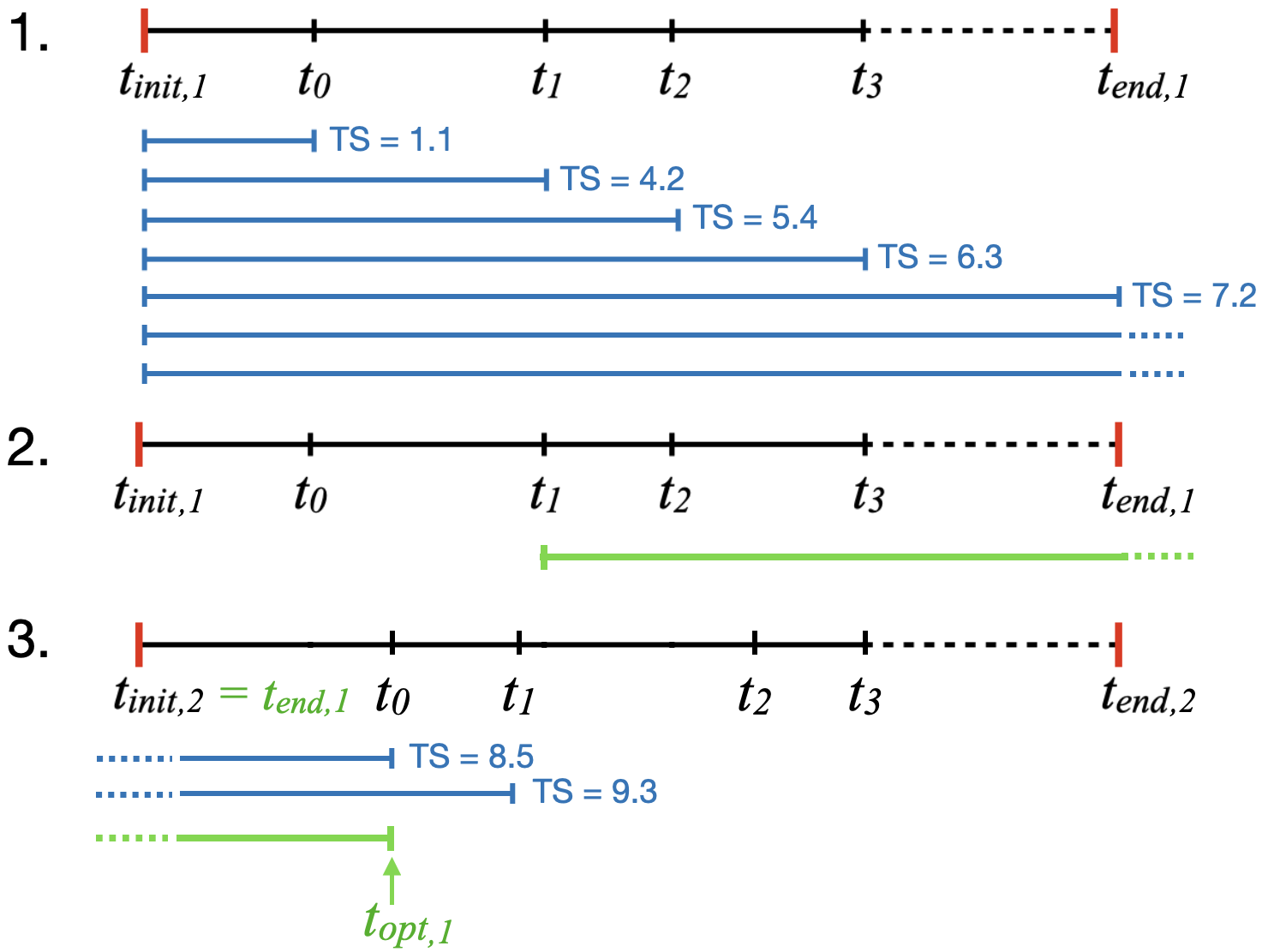}
    \caption{Visualisation of the steps of the time bin search algorithm in the case of an unfinished range of event timestamps whose TS are within the target range (here [4,9]). 
    }
    \label{fig:tbins_ext}
    \vspace{-0.2cm}
\end{figure}

Since the search is performed chronologically, if one takes the first event's timestamp within a time window that produces a TS within the target range as $t_{opt}$, exceedingly short time bins tend to occur. If one runs these time bins through the Fermi-LAT analysis, they will be evaluated as sudden strong fluxes with sudden hard spectra. For this reason, we take the last timestamp with a TS that is still within the target range instead of the first. If the last timestamp of a time window still produces a TS within the target range, then the time window is increased (while keeping the same initial $t_{init}$) until timestamps which produce TS above the target range are found.
This is depicted in Figure \ref{fig:tbins_ext}.

The final time bins can then be fed through the \textit{Fermipy} python API to produce the final light curve with adapted time-binning.

\section{Example light curves}

In this section, we provide a few example light curves that we produced using \emph{flashcurve} and compare them with the light curves based on the method of \citet{Lott_2012}\footnote{Provided by Narek Sahakyan}. Four independent test sources were studied:
\begin{itemize}
    \item TXS 0506+056, an IBL/LBL blazar which was identified as a source of high-energy neutrinos \citep{10.1093/mnras/sty1852}
    \item CTA 102 a high-frequency peaked BL Lac (HBL) 
    \item MKN 421, an LBL, which is one of the closest and \\brightest blazars 
    \item MKN 501, a blazar with relatively lower gamma-ray flux.
\end{itemize}

For the generation of the \emph{flashcurve} light curves, we use the entire energy range between 100 MeV and 300 GeV. The light curves produced with the classic approach, on the contrary, are made with an energy cut-off based on the optimal minimal energy $E_{\text{min}}$ calculated with the \citet{Lott_2012} methodology. 

Note that the adaptive binning light curves made with the \citet{Lott_2012} method were produced with a constant relative flux error, as the constant TS mode was not available at the time of this study\footnote{\sloppy\url{https://www.slac.stanford.edu/~lott/tuto_adaptive_binning.pdf}, last accessed: 01.07.2024}. In order to achieve comparable results, the range of TS calculated from the constant relative flux error method was set as the thresholds for \emph{flashcurve}. The details are listed in Table \ref{tab:src_ts}.

\begin{table}[H]
\fontsize{8.5}{11}\selectfont
\centering
\begin{tabular}{|l|l|l|}
\hline
\textbf{Source} & \textbf{Min. TS} & \textbf{Max. TS} \\ \hline
TXS 0506+056 & 25 & 50  \\ \hline
CTA 102 & 50 & 75  \\ \hline
MKN 421 & 115 & 170 \\ \hline
MKN 501 & 75 & 125 \\ \hline

\end{tabular}
\caption{List of test sources used to make example light curves, along with the corresponding target range (Min. TS to Max. TS) for the time bin search algorithm.}
\label{tab:src_ts}
\end{table}
\begin{figure*}[h!]
    \centering
    \begin{minipage}[b]{0.495\textwidth}
        \centering
        \includegraphics[scale=0.33]{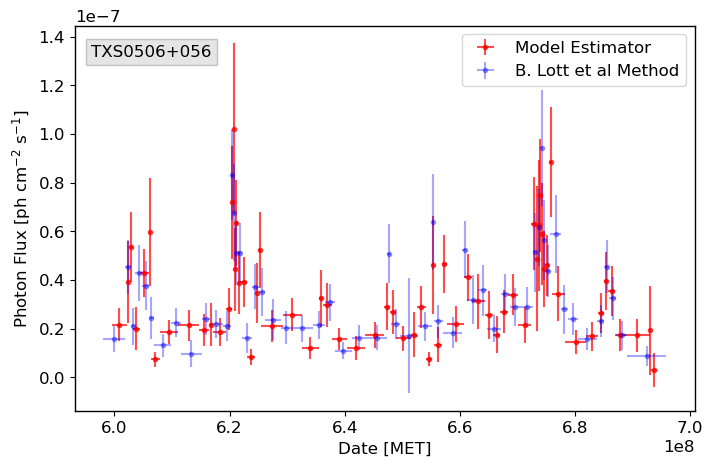}
    \end{minipage}
    \hfill
    \begin{minipage}[b]{0.495\textwidth}
        \centering
        \includegraphics[scale=0.335]{
        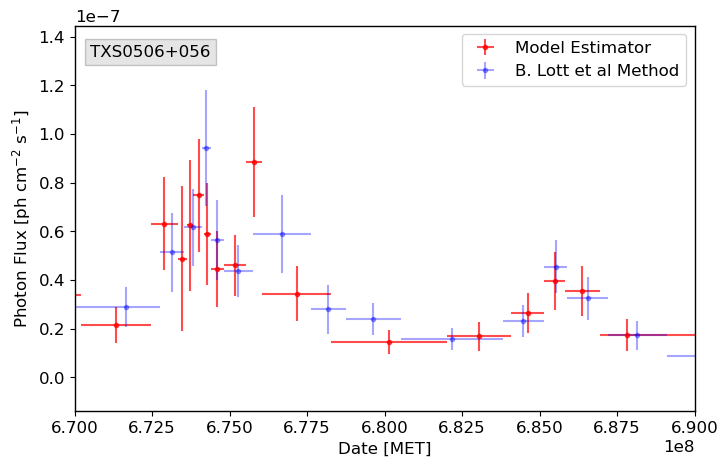
        %cta_lc_full.png
        }
        % \subcaption{Plot 2}\label{fig:plot2}
    \end{minipage}
    \vfill
    \begin{minipage}[b]{0.495\textwidth}
        \centering
        \includegraphics[scale=0.33]{
        %mkn421_lc_full.png
        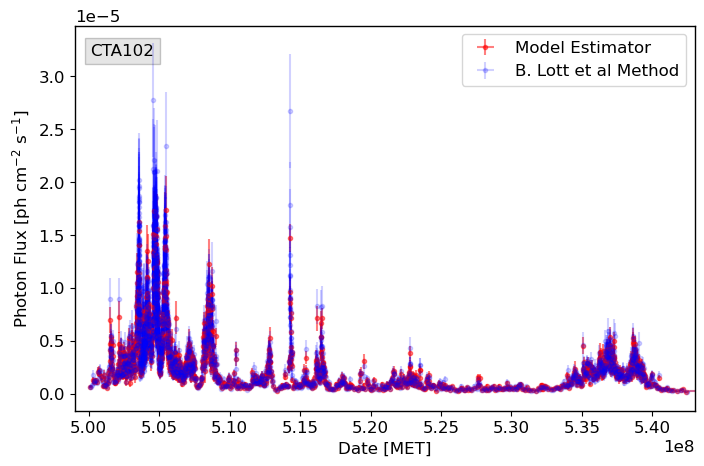
        }
        % \subcaption{Plot 3}\label{fig:plot3}
    \end{minipage}
    \hfill
    \begin{minipage}[b]{0.495\textwidth}
        \centering
        \includegraphics[scale=0.338]{
        %mkn501_lc_full.png
        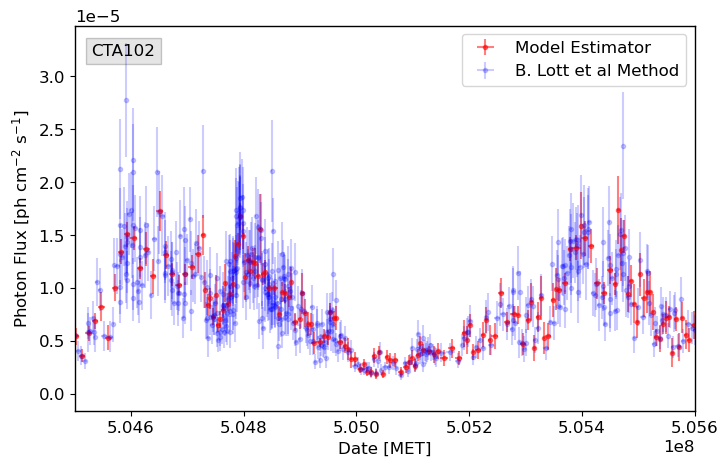
        }
        % \subcaption{Plot 4}\label{fig:plot4}
    \end{minipage}
    \vfill
        \begin{minipage}[b]{0.495\textwidth}
        \centering
        \includegraphics[scale=0.33]{
    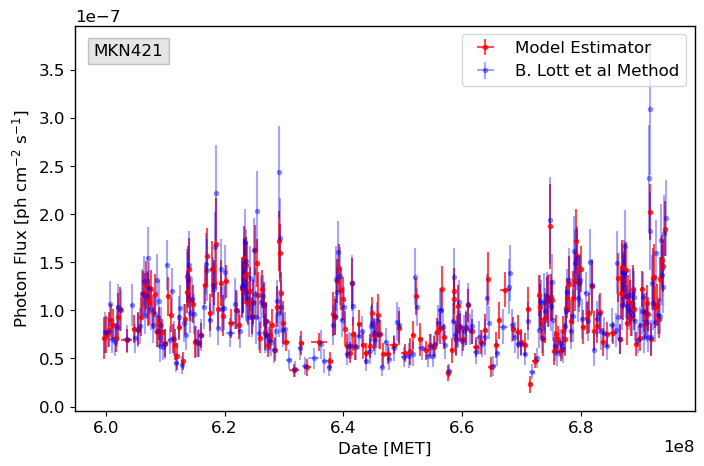
    %txs_lc_part1.png
        }
    \end{minipage}
    \hfill
    \begin{minipage}[b]{0.495\textwidth}
        \centering
        \includegraphics[scale=0.335]{
        %cta_lc_part1.png
        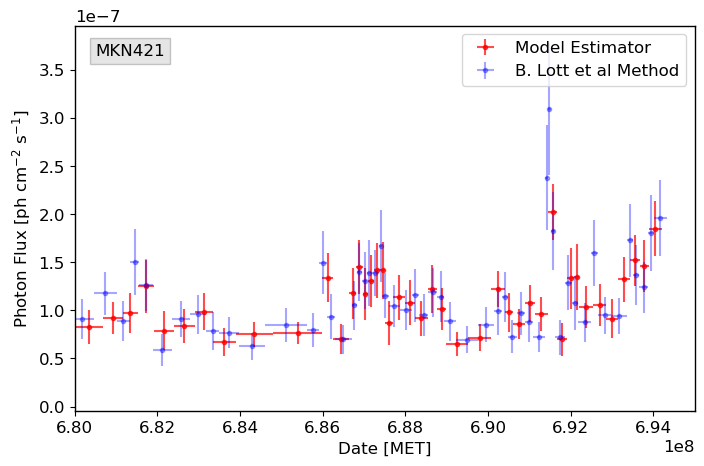
        }
        % \subcaption{Plot 2}\label{fig:plot2}
    \end{minipage}
    \vfill
    \begin{minipage}[b]{0.495\textwidth}
        \centering
        \includegraphics[scale=0.33]{
        %mkn421_lc_part1.png
        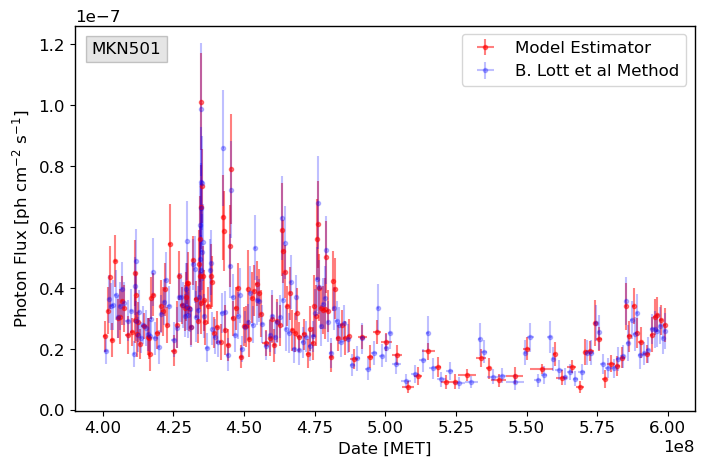
        }
        % \subcaption{Plot 3}\label{fig:plot3}
    \end{minipage}
    \hfill
    \begin{minipage}[b]{0.495\textwidth}
        \centering
        \includegraphics[scale=0.338]{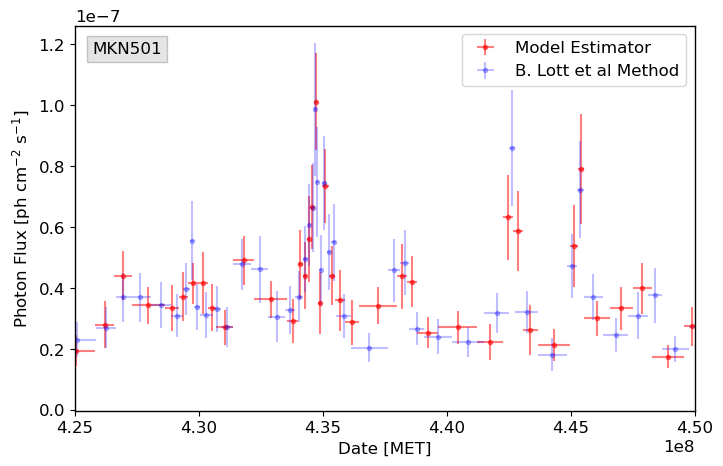}
        % \subcaption{Plot 4}\label{fig:plot4}
    \end{minipage}
    \caption{Light curves of four test sources comparing  \emph{flashcurve} with the method of \citet{Lott_2012}. The four right plots are a zoomed-in section of interest of each source, with flux error and time bin length as vertical and horizontal errors, respectively.}
    \label{fig:main_lc_collage}
\end{figure*}

\begin{figure*}[h!]
    \centering
    \begin{minipage}[b]{0.495\textwidth}
        \centering
        \includegraphics[scale=0.342]{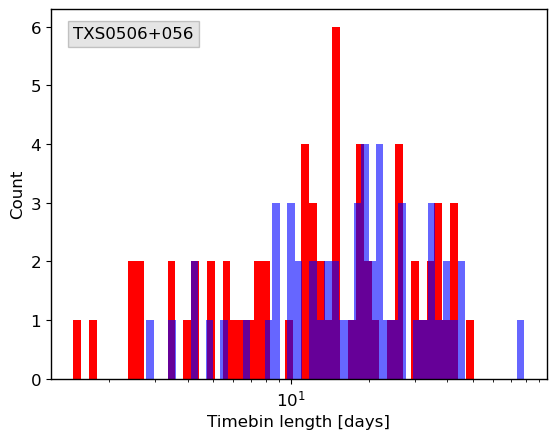}
    \end{minipage}
    \hfill
    \begin{minipage}[b]{0.495\textwidth}
        \centering
        \includegraphics[scale=0.345]{
        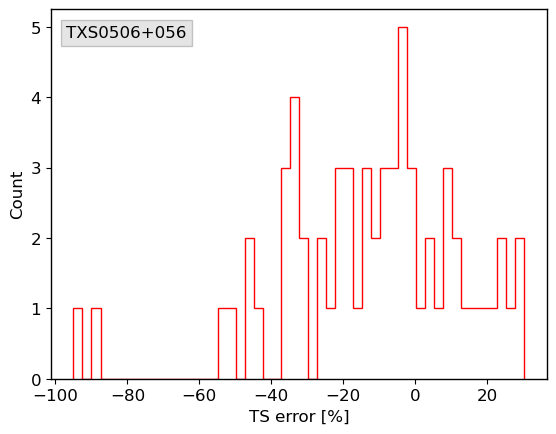
        %cta_tb_dist_full.png
        }
        % \subcaption{Plot 2}\label{fig:plot2}
    \end{minipage}
    \vfill
    \begin{minipage}[b]{0.495\textwidth}
        \centering
        \includegraphics[scale=0.342]{
        %mkn421_tb_dist_full.png
        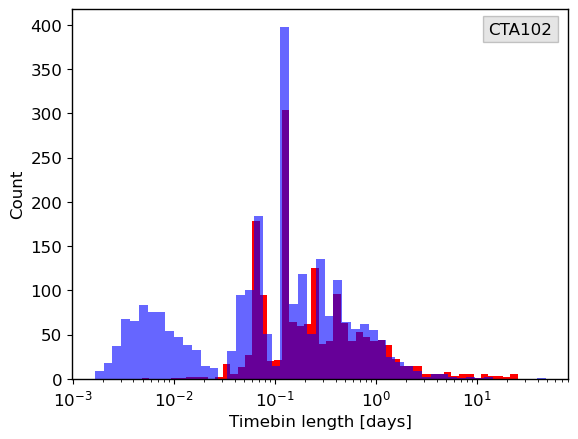}
        % \subcaption{Plot 3}\label{fig:plot3}
    \end{minipage}
    \hfill
    \begin{minipage}[b]{0.495\textwidth}
        \centering
        \includegraphics[scale=0.347]{
        %mkn501_tb_dist_full.png
        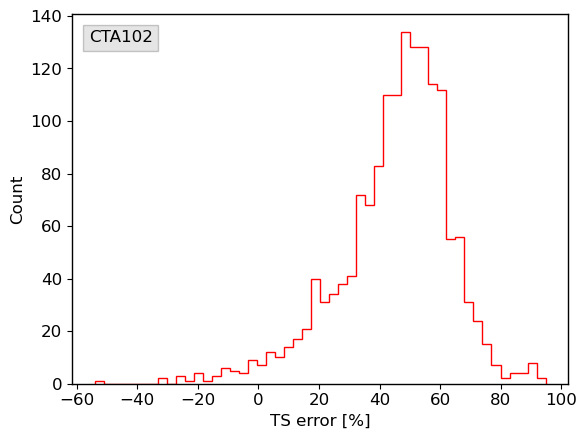
        }
        % \subcaption{Plot 4}\label{fig:plot4}
    \end{minipage}
    \vfill
        \begin{minipage}[b]{0.495\textwidth}
        \centering
        \includegraphics[scale=0.342]{
        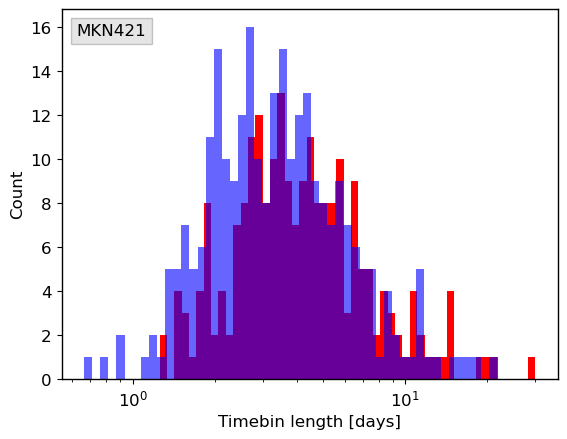
        %txs_tserr_full.png
        }
    \end{minipage}
    \hfill
    \begin{minipage}[b]{0.495\textwidth}
        \centering
        \includegraphics[scale=0.345]{
        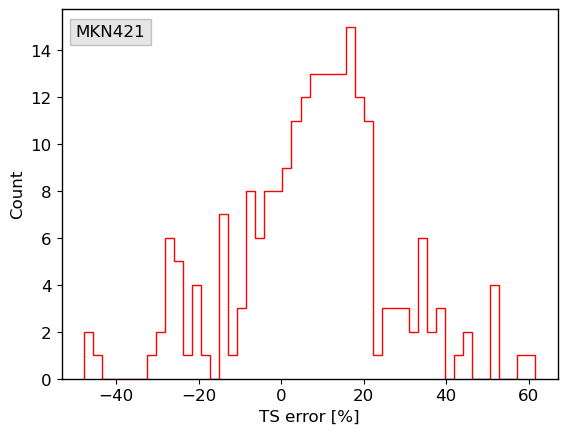
        %cta_tserr_full.png
        }
        % \subcaption{Plot 2}\label{fig:plot2}
    \end{minipage}
    \vfill
    \begin{minipage}[b]{0.495\textwidth}
        \centering
        \includegraphics[scale=0.342]{
        %mkn421_tserr_full.png
        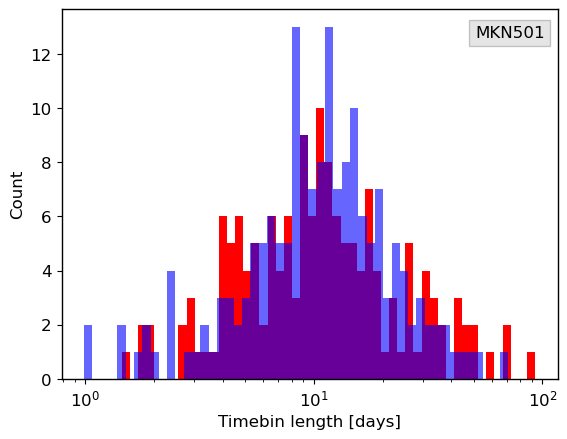}
        % \subcaption{Plot 3}\label{fig:plot3}
    \end{minipage}
    \hfill
    \begin{minipage}[b]{0.495\textwidth}
        \centering
        \includegraphics[scale=0.347]{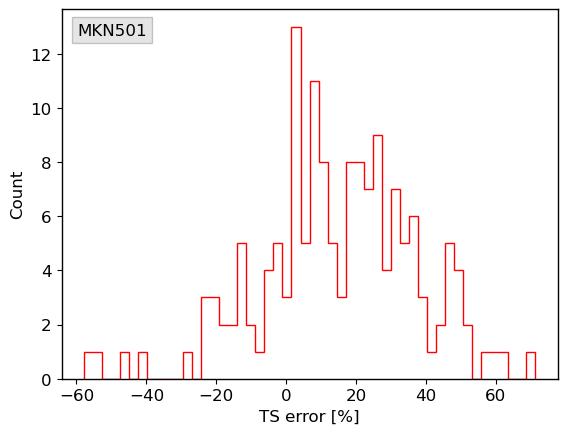}
        % \subcaption{Plot 4}\label{fig:plot4}
    \end{minipage}
    \caption{\emph{Left plots}: Histograms of the time bin lengths using \emph{flashcurve} (red) and the method of \citet{Lott_2012} (blue) for the four test sources.\\
    \emph{Right plots}: Histograms of the relative TS deviation of the TS estimated using \emph{flashcurve} from the TS calculated using the full Fermi-LAT likelihood analysis for the four test sources.}
    \label{fig:tb_tserr_collage}
\end{figure*}

\begin{figure*}[h!]
    \centering
    \begin{minipage}[b]{0.495\textwidth}
        \centering
        \includegraphics[scale=0.343]{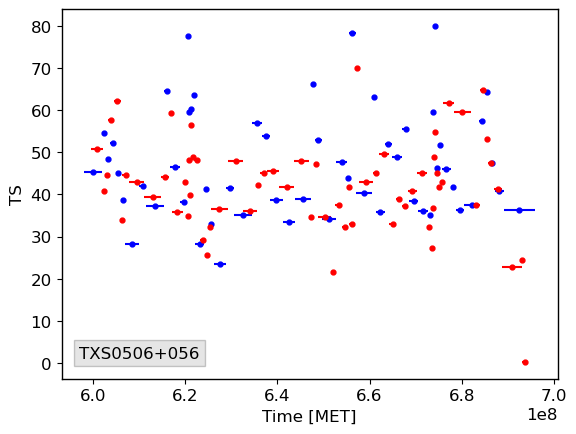}
    \end{minipage}
    \hfill
    \begin{minipage}[b]{0.495\textwidth}
        \centering
        \includegraphics[scale=0.345]{
        %cta_ts_full.png
        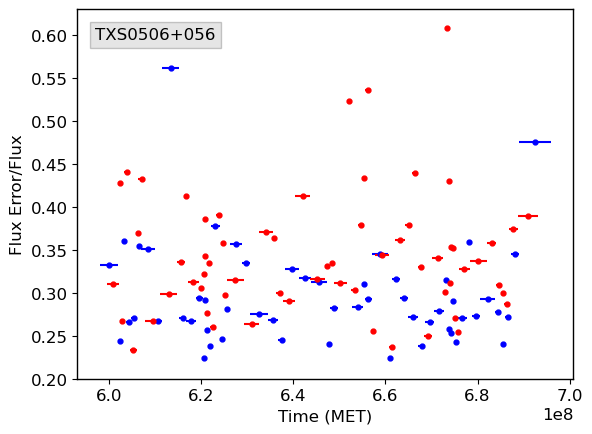
        }
        % \subcaption{Plot 2}\label{fig:plot2}
    \end{minipage}
    \vfill
    \begin{minipage}[b]{0.495\textwidth}
        \centering
        \includegraphics[scale=0.343]{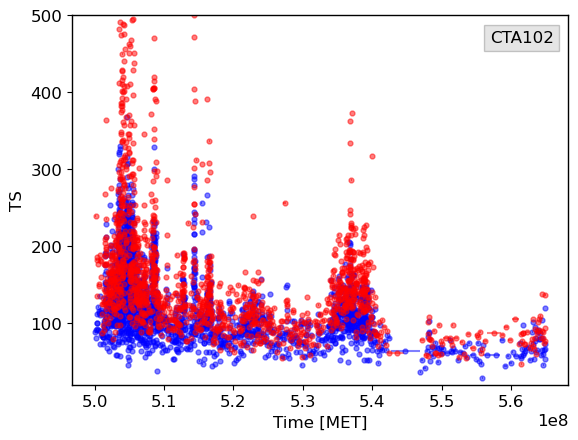
        %mkn421_ts_full.png
        }
        % \subcaption{Plot 3}\label{fig:plot3}
    \end{minipage}
    \hfill
    \begin{minipage}[b]{0.495\textwidth}
        \centering
        \includegraphics[scale=0.348]{
        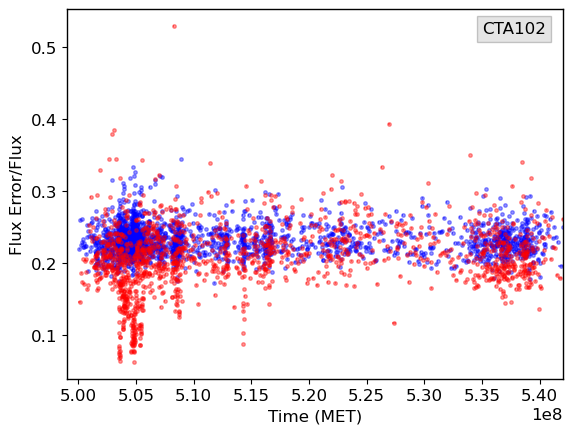
        %mkn501_ts_full.png
        }
        % \subcaption{Plot 4}\label{fig:plot4}
    \end{minipage}
    \vfill
        \begin{minipage}[b]{0.495\textwidth}
        \centering
        \includegraphics[scale=0.343]{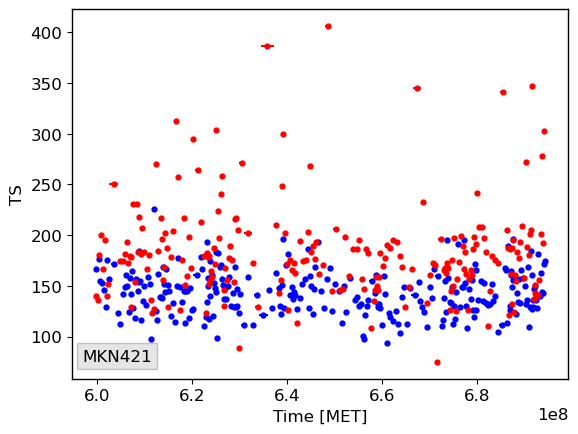
        %txs_fe_full.png
        }
    \end{minipage}
    \hfill
    \begin{minipage}[b]{0.495\textwidth}
        \centering
        \includegraphics[scale=0.345]{
        %cta_fe_full.png
        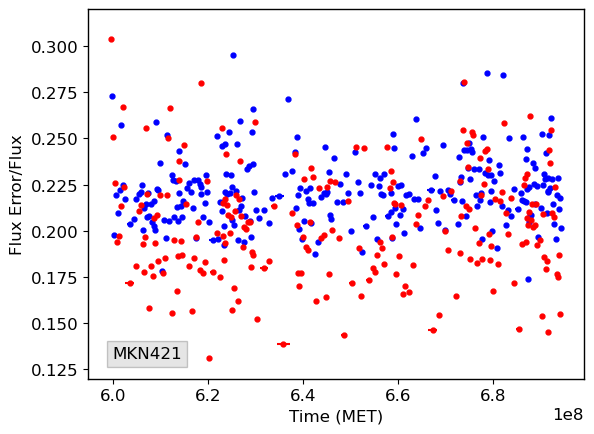
        }
        % \subcaption{Plot 2}\label{fig:plot2}
    \end{minipage}
    \vfill
    \begin{minipage}[b]{0.495\textwidth}
        \centering
        \includegraphics[scale=0.343]{
        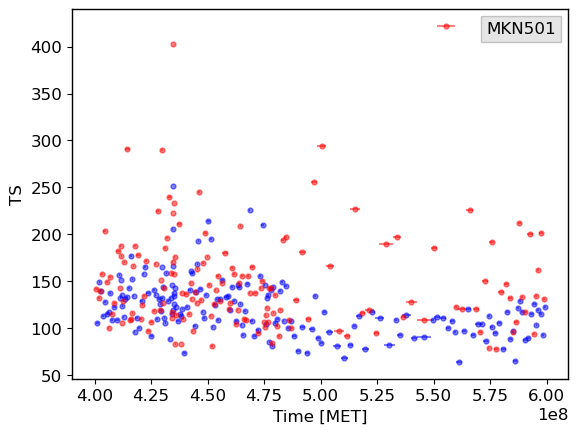
        %mkn421_fe_full.png
        }
        % \subcaption{Plot 3}\label{fig:plot3}
    \end{minipage}
    \hfill
    \begin{minipage}[b]{0.495\textwidth}
        \centering
        \includegraphics[scale=0.348]{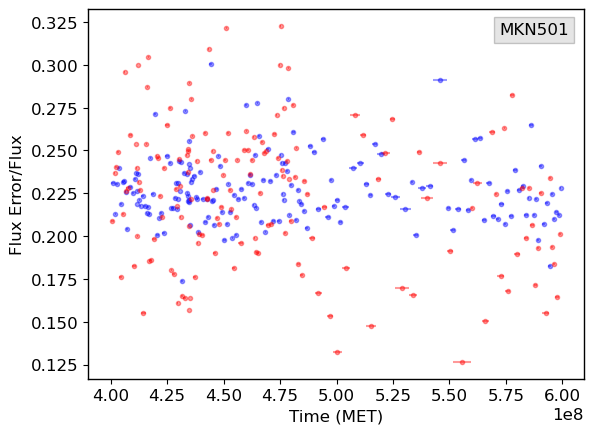}
        % \subcaption{Plot 4}\label{fig:plot4}
    \end{minipage}
    \caption{\emph{Left plots}: TS calculated using the Fermi-LAT analysis for the time bins found using our \emph{flashcurve}(red) and the method of \citet{Lott_2012} (blue) for the four test sources, with respective time bin lengths. In the case of CTA 102, a few extremely high TS are cut out for better visualisation. \\
    \emph{Right plots}: Flux error versus flux ratios for the time bins found by the two methods. }
    \label{fig:ts_fe_collage}
\end{figure*}
\begin{figure*}[t]
    \vspace*{-0.1cm}
    \centering
    \begin{minipage}{0.495\textwidth}
        \centering
        \includegraphics[scale=0.33]{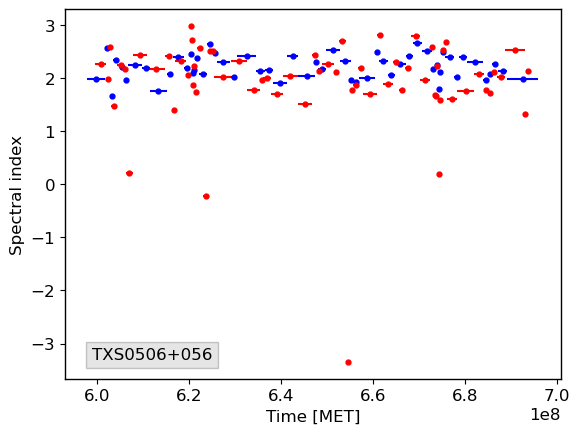}
    \end{minipage}
    \hfill
    \begin{minipage}{0.495\textwidth}
        \centering
        \includegraphics[scale=0.335]{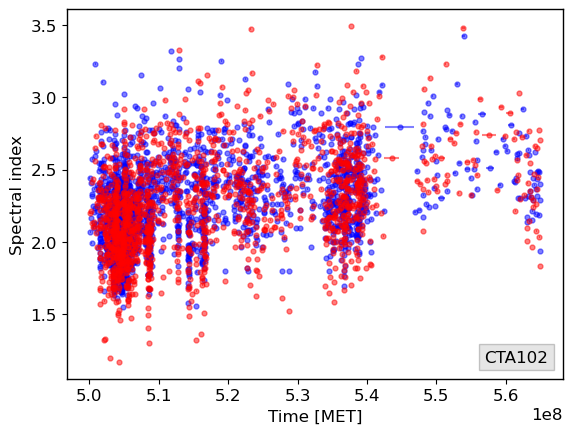}
        % \subcaption{Plot 2}\label{fig:plot2}
    \end{minipage}
    \vfill
    \begin{minipage}{0.495\textwidth}
        \centering
        \includegraphics[scale=0.33]{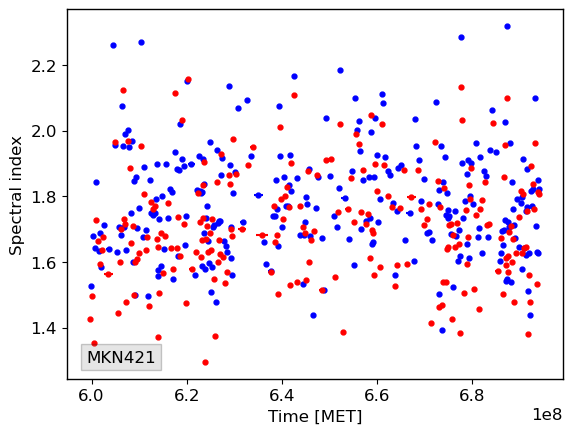}
        % \subcaption{Plot 3}\label{fig:plot3}
    \end{minipage}
    \hfill
    \begin{minipage}{0.495\textwidth}
        \centering
        \includegraphics[scale=0.338]{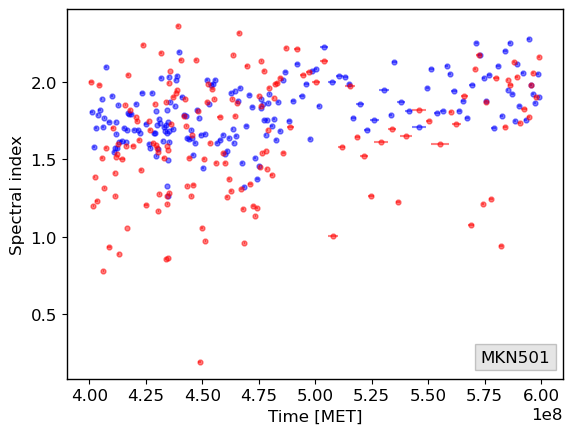}
        % \subcaption{Plot 4}\label{fig:plot4}
    \end{minipage}
    \caption{Best-fit spectral indices for the time bins found using \emph{flashcurve}(red) and the method of \citet{Lott_2012} (blue) for the four test sources.}
    \label{fig:ind_lc_collage}
\end{figure*}

The results of our study can be seen in Figures \ref{fig:main_lc_collage} to \ref{fig:ind_lc_collage}. At a general glance, \emph{flashcurve} performs well, producing light curves with relatively constant significance time bins and without upper limits while also capturing the finer dynamics of the source flux. Moreover, we see reasonable agreement with the \citet{Lott_2012} method. 

Figure \ref{fig:main_lc_collage} shows the light curves of each of the test sources. We observe a good agreement with the classic method for three of the selected sources. This is the case for TXS 0506, MKN 421, and MKN 501. Looking at the histograms of the time bin lengths in Figure \ref{fig:tb_tserr_collage} and the spectral indices in Figure \ref{fig:ind_lc_collage} of these three sources, it is evident that the two methods produce similar results, except for a few short time bins producing overly hard spectra. This is due to \emph{flashcurve}'s time bin search algorithm, which only produces time bins that begin and end on photon event time stamps.

In the next step, we study the TS and relative flux errors of the light curves in Figure \ref{fig:ts_fe_collage} as well as the relative deviation of the estimated TS from the Fermi-LAT analysis in Figure \ref{fig:tb_tserr_collage}. Since \emph{flashcurve}'s time bins for TXS 0506+056 were relatively shorter, this resulted in higher relative flux errors and lower TS than that of the \citet{Lott_2012} method. Overall, however, we observe consistency between the two methods. We also note that \emph{flashcurve} generally produces constant TS across all its time bins. Similarly, there is good consistency between the methods for MKN 421 and MKN 501. However, in this case, \emph{flashcurve} tends to predict larger time bins, resulting in relatively larger TS with smaller flux error. 
This could be attributed to the energy and proximity thresholds of the time bin search algorithm, which filtered out intermediate photon events. Again, we see a relatively constant and decently accurate TS prediction overall. 

The situation is different for CTA 102. Here, the two methods produce different results. CTA 102 was chosen due to its high activity and hard fluxes \citep{refId0}, which was meant to provide a challenging case for both light curve algorithms. Compared to the true TS, it appears that \emph{flashcurve} frequently under-predicts the TS value. This also results in small relative flux errors in the light curve, especially in its highly active flaring periods. In these periods, we also see larger variance in true TS; however, looking at relative TS error, we see a clear peak indicating that TS prediction, although collectively deviating from the target range, was consistent across all time bins. This, again, shows that the varied TS was more a fault of \emph{flashcurve}'s time bin search algorithm rather than its TS prediction accuracy. Also, we see that the method of \citet{Lott_2012} also displayed a relatively larger variance in TS and relative flux error during these periods, showing that these periods are generally difficult to process for either method.

Both \emph{flashcurve} and the classic method have similar time bin length distributions. However, there is an additional population of extremely short time-bins, which was found by the \citet{Lott_2012} method, as seen in CTA 102's time bin length distribution in Figure \ref{fig:tb_tserr_collage}. 
These shorter time bins do not necessarily indicate a better choice, as many of them come from the period of larger variance in flux error for the method of \citet{Lott_2012}. 
Using energy and proximity thresholds in \emph{flashcurve}'s time bin search algorithm can cause it to predict fewer extremely short time bins. 

In less active periods, we see more constant TS and similar flux dynamics relative to the classic method, as seen in the additional zoomed-in light curve of CTA 102 in Figure \ref{fig:extra_lc_collage}.

We also see that, again, no upper limits were produced by the time bins predicted by \emph{flashcurve}.

A final remark about the difference in the estimation of the time bins between the two methods is that it was clear that \emph{flashcurve} is much faster than any previous method. The time-scales for producing these time bins (not including the Fermi-LAT analysis after that) ranged from $\sim$30 min to a few hours in the case of the very high activity sources. Using the method of \citet{Lott_2012} can take up to several days in comparison. 
This was mainly made possible using a neural network model that produces fast TS estimations. This can be sped up by multiprocessing, with around 100 threads used in these example cases. The main bottleneck of \emph{flashcurve} is its time bin search algorithm, which can be replaced with a quicker and more sophisticated algorithm in the future. This might also improve the light curves by allowing the energy and proximity thresholds to be set lower in order to avoid missing shorter duration time bins, as in the case of CTA 102.

\newpage
\section{Conclusion and Outlook}

Using a machine-learning-based estimator model, we have presented a novel approach to adaptive binning of Fermi-LAT gamma-ray light curves. By leveraging neural networks, specifically convolutional neural networks, \emph{flashcurve} estimates the test statistic for gamma-ray sources with respect to the pure background for gamma-ray events detected in a given time window. 

The method offers a significant improvement in computational speed over traditional methods like \citet{Lott_2012}. 
Light curves generated with \emph{flashcurve} aim for a constant significance. This allows for fast variations in flux to be captured while at the same time avoiding upper limits in quiet periods. We validated \emph{flashcurve} against multiple test gamma-ray sources. The results showed that for most sources, \emph{flashcurve} produces high-quality light curves, which contain time bins with relatively constant TS, have no upper limits, and encompass smaller flux variations whenever appropriate. Results were also generally similar to that of the \citet{Lott_2012} method, with only minor differences observed, particularly in highly active sources like CTA 102. Both methods, however, generally produced results with higher variance in relative flux error. In less active periods, however, \emph{flashcurve}'s performance still demonstrated accurate results.

\begin{figure*}[h!]
    \centering
    \vspace*{-0.45cm}
        \begin{minipage}{0.495\textwidth}
        \centering
        \includegraphics[scale=0.31]{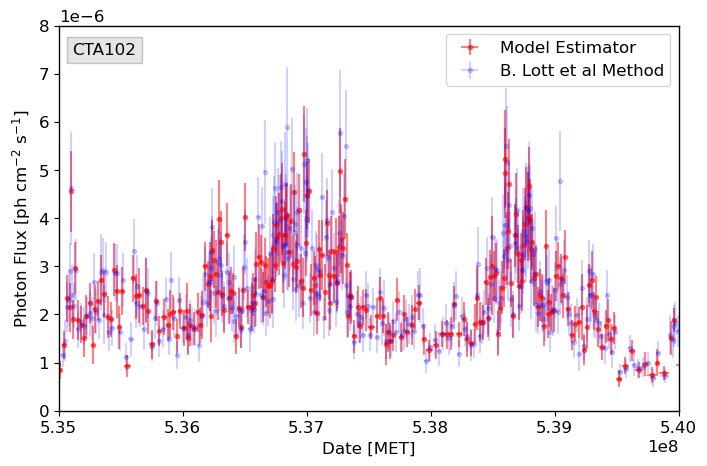}
    \end{minipage}
    \hfill
    \begin{minipage}{0.495\textwidth}
        \centering
        \includegraphics[scale=0.31]{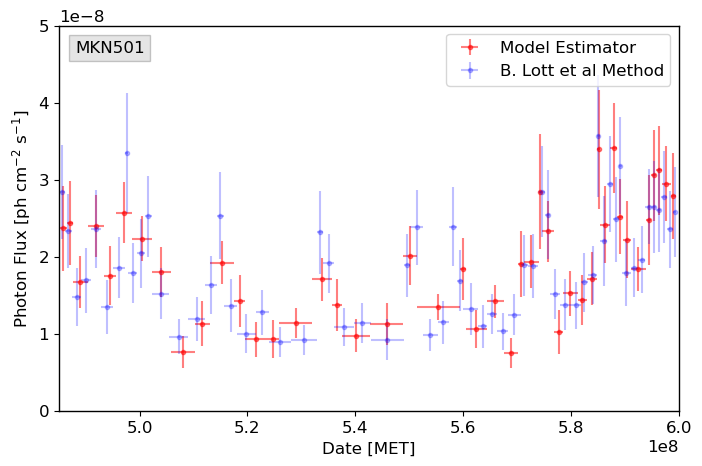}
        % \subcaption{Plot 2}\label{fig:plot2}
    \end{minipage}
    \caption{
    Continuation of Figure \ref{fig:main_lc_collage}. Additional zoomed-in light curves.}
    \label{fig:extra_lc_collage}
    \vspace*{-0.5cm}
\end{figure*}

The training process, involving over 1.5 million time bins, utilised a residual block-based convolutional neural network architecture optimised with the ADAM optimiser. Despite challenges such as data imbalance towards lower TS values, the model for \emph{flashcurve} generalised well on unseen test sets. Validation against traditional methods showcased the potential of this approach in producing reliable light curves for gamma-ray sources at quick speeds, emphasising the importance of future improvements in model training and data handling techniques.

Future development of \emph{flashcurve} will focus on several key areas. Firstly, expanding the dataset to include more sources, particularly those near the Galactic plane and extended sources, will help the generalisation capabilities of the estimator. Additional input layers could help to include information on the (known) sources in the region. Secondly, similar to the prediction of the TS values, the same approach can also be used to predict the flux uncertainty as an alternative method for generating adaptive binning light curves. 

The model can be continuously improved by using \\\emph{flashcurve}'s output as samples for the training. This way, the network could be trained in an infinite training loop for maximal accuracy, only limited by computational resources. 

As for the time bin search algorithm, other numerical approaches, such as a bisection method (similar to the one described by \citet{378033}) could, in future, be implemented to help further speed up computation.
This would also allow testing a more extensive range of time windows, reducing the variance in the light curve test statistics or flux uncertainties.

In general, using a machine-learning estimator to retrieve the time windows of adaptive-binning light curves is not limited to Fermi-LAT data. It could be similarly used to produce light curves in any other wavelength if sufficient training data is available. 

\section{Acknowledgements}
\vspace{-0.2cm}
We thank Martina Karl for proofreading the paper, Paolo Giommi and Narek Sahakyan for their consultation and provision of adaptive binning light curves, and Elisa Resconi for the green-lighting of this project. Funding for this research was provided through the SFB1258 grant. We also thank the unknown reviewers for reading the paper thoroughly and providing valuable comments. 

\vspace{-0.35cm}
\bibliographystyle{elsarticle-harv} 
\bibliography{cas-refs}

\end{document}